\font\tenfrakturb=eufb10
\font\tenfraktur=eufm10
\font\tenmsbm=msbm10
\font\sevenfrakturb=eufb7
\font\sevenfraktur=eufm7
\font\sevenmsbm=msbm7
\font\fivefrakturb=eufb5
\font\fivefraktur=eufm5
\font\fivemsbm=msbm5
\newfam\bgothicfam
\newfam\gothicfam
\newfam\msbmfam
\textfont\bgothicfam = \tenfrakturb \scriptfont\bgothicfam=\sevenfrakturb
\scriptscriptfont\bgothicfam=\fivefrakturb

\textfont\gothicfam = \tenfraktur \scriptfont\gothicfam=\sevenfraktur
\scriptscriptfont\gothicfam=\fivefraktur

\textfont\msbmfam = \tenmsbm \scriptfont\msbmfam=\sevenmsbm
\scriptscriptfont\msbmfam=\fivemsbm

\def\Bbb{\tenmsbm\fam\msbmfam}

\def\goth{\tenfraktur\fam\gothicfam}
\catcode`@=11
\def\renewcounter#1{\@definecounter{#1}\@ifnextchar[{\@newctr{#1}}{}}
\documentstyle[12pt]{article}
\renewcounter{equation}[section]

\def\sector#1#2{\ {\scriptstyle #1}\hskip 1mm
\mathop{\fbox{\rule[.3cm]{.4cm}{0cm}$\frac{}{}$}}\limits_{\lower 1mm
\hbox{$\scriptstyle#2$}}
\hskip 1mm}

\begin{document}
\title{Asymptotics of Elliptic Genera of Symmetric Products and Dyonic 
Black Hole Degeneracy}
\author{A.A. Bytsenko \thanks{E-mail: abyts@fisica.uel.br\,\,\,\,\,
On leave from Sankt-Petersburg State Technical University, Russia},\,\,\,
A.E. Gon\c calves \thanks{E-mail: goncalve@fisica.uel.br}\\
Departamento de Fisica, Universidade Estadual de Londrina,\\
Caixa Postal 6001, Londrina-Parana, Brazil\\
and\\
S.D. Odintsov \thanks{E-mail: odintsov@quantum.univalle.edu.co\,\,\,\,\,
On leave from Tomsk Pedagogical University, Russia}\\
Departamento de Fisica, Universidad del Valle,\\
AA 25360, Cali, Colombia}

\maketitle
\begin{abstract}

We calculate asymptotic expansions of elliptic genera for a supersymmetric
sigma model on the $N-$ fold symmetric product $S^NM$ of a K\"{a}hler manifold
$M$ and for $N=2$ superconformal field theory. Asymptotic expansions for the
degeneracy of dyonic black hole spectrum are also derived.

\end{abstract}

{\bf I}.\,\,\,\,\,  In this note we use an identity proved in 
\cite{dijk97-185-197} that equates elliptic genus partition function of a 
supersymmetric sigma model on the $N$-fold symmetric product $S^NM$ of 
$M$\,( $S^NM=M^N/S_N$, $S_N$ is the symmetric group 
of $N$ elements) to the partition function of a second quantized string theory
on the space $S^1\otimes M$. The generating function of these elliptic
genera is almost an authomorphic form for $O(3,2,{\Bbb Z})$\, $T-$ duality 
group. We shall derive asymptotic expansion of elliptic genus as well as
asymptotic growth for the degeneracy of dyonic spectrum in four-dimensional
$N=4$ string theory. It is known that the counting formula is manifestly 
symmetric under the duality group; we wish to present  its 
asymptotic growth in explicit form,
 which reproduces the macroscopic Bekenstein-Hawking 
entropy.  

We start with a supersymmetric sigma model on $S^NM$. The Hilbert space of an 
orbifold field theory can be decomposed
into twisted sectors ${\cal H}_{\gamma}$, that are labelled by the conjugacy
classes $\{\gamma\}$ of the orbifold group $S_N$
\cite{dixo85-261-620,dixo86-274-285,dijk97-185-197}. For given twisted sector 
one can keep the states invariant under the centralizer subgroup 
$\Gamma_{\gamma}$
related to the element $\gamma$. Let ${\cal H}_{\gamma}^{\Gamma_{\gamma}}$ is 
an invariant
subspace associated with $\Gamma_{\gamma}$; the total orbifold Hilbert space 
takes 
the form ${\cal H}(S^NM)=\bigoplus_{\{\gamma\}}{\cal H}_{\gamma}^{\Gamma_
{\gamma}}$.
Taking into account the group $S_N$ one can compute the conjugacy classes
$\{\gamma\}$ by set of partitions $\{N_n\}$ of $N$, namely $\sum_nnN_n=N$, 
where 
$N_n$ is the multiplicity of the cyclic permutation $(n)$ of $n$ elements in
the decomposition of $\gamma$: $\{\gamma\}=\sum_{j=1}^s(j)^{N_j}$. For this
conjugacy class the centralizer subgroup of a permutation $\gamma$ is
$\Gamma_{\gamma}=S_{N_1}\bigotimes_{j=2}^s\left(S_{N_j}>\!\!\!\lhd 
{\Bbb Z}_j^{N_j}\right)$ \cite{dijk97-185-197}, where each subfactor 
$S_{N_n}$ and ${\Bbb Z}_n$ 
permutes the $N_n$ cycles $(n)$ and acts within one cycle $(n)$ 
correspondingly.
Following the lines of Ref. \cite{dijk97-185-197} we may decompose each 
twisted sector ${\cal H}_{\gamma}^{\Gamma_{\gamma}}$ into the product over the 
subfactors $(n)$ of $N_n$-fold symmetric tensor products,
${\cal H}_{\gamma}^{\Gamma_{\gamma}}=\bigotimes_{n>0}S^{N_n}{\cal H}_{(n)}^
{{\Bbb Z}_{n}}$, 
where $S^N{\cal H}\equiv (\bigotimes^N{\cal H})^{S_N}$. 
Let $\chi({\cal H};q,y)$ be the partition function for every (sub) Hilbert 
space of a supersymmetric sigma model; $q={\bf e}[\tau],\,\,
y={\bf e}[z]$, and ${\bf e}[x]\equiv\exp[2\pi ix]$. 
It can be shown that the partition function coincides with the elliptic genus
\cite{sche86-177-317,sche87-287-317,witt87-109-525,land88,eguc89-315-193,
kawa94-414-191}.

In general the genus is a homomorphism: $\Omega\mapsto\Bbb C$, where $\Omega$ 
is the Thom ring of oriented cobordisms. The following relations for the 
genus hold:
$$
\chi(M_1\bigsqcup M_2;q,y) = \chi(M_1;q,y)+ \chi(M_2;q,y)
\mbox{,}   
\eqno{(1)}
$$
$$
\chi(M_1\bigotimes M_2;q,y) = \chi(M_1;q,y)\cdot \chi(M_2;q,y)
\mbox{,}   
\eqno{(2)}
$$
$$
\chi(M_1;q,y)=\chi(M_2;q,y)\,\,\, \mbox{if}\,\,\,\,M_1\,\,\,\mbox{and}
\,\,\,M_2 \,\,\,\,\,\mbox{are cobordant}
\mbox{.}
\eqno{(3)}
$$
The elliptic genus admits the following identities:
$$
\chi({\cal H}\bigoplus {\cal H}^\prime;q,y) = \chi({\cal H};q,y)+    
\chi({\cal H}^\prime;q,y)
\mbox{,}   
\eqno{(4)}
$$
$$        
\chi({\cal H}\bigotimes {\cal H}^\prime;q,y)=\chi({\cal H};q,y)\cdot
\chi({\cal H}^\prime;q,y)
\mbox{.}
\eqno{(5)}
$$
For any vector bundle $V$ over a smooth manifold we can define 
formal sums
$$
{\bigwedge}_qV=\bigoplus_{k\geq 0}^dq^k{\bigwedge}^kV,\,\,\,
S_qV=\bigoplus_{k\geq 0}^{\infty}q^kS^kV,\,\,\,
\chi_y(T_M)=\bigoplus_{s=0}^{d}(-y)^s\chi\left({\bigwedge}^sT_M\right)
\mbox{,}
\eqno{(6)}
$$
where $T_{M}$ is the holomorphic tangent bundle of K\"{a}hler manifold $M$
of complex dimension $d$,\, 
$\bigwedge^k\,\,\,\,(S^k)$ denote the $k$-th exterior (symmetric 
product).
Note that for a Calabi-Yau space the $\chi_y-$ genus \cite{hirz78} is a weak 
Jacobi form of weight zero and index $d/2$ and it transforms as
$\chi_y(T_M)=(-1)^{r-d}y^r\chi_{y^{-1}}(T_M)$.
This relation can also be derived from the Serre duality 
$H^j(M;{\bigwedge}^sT_M)\cong H^{d-j}(M;{\bigwedge}^{r-s}T_M)$. 

The Riemann-Roch-Hirzebruch theorem: 
$$                                                                            
\chi(M;q,y)\stackrel{def}=\sum (-1)^j\mbox{dim}H^j(M;E_{q,y})=
\int_{M}\mbox{ch}(E_{q,y})Td(M)
\mbox{,}
\eqno{(7)}
$$
gives an equivalent definition of the elliptic genus.
The bundle $E_{q,y}$ of a characteristic class is
$$
E_{q,y}=y^{-\frac{d}{2}}\bigotimes_{n\geq 1}^{\infty}\left(
{\bigwedge}_{-yq^{n-1}}
T_{M}\bigotimes_{n\geq 1}^{\infty}{\bigwedge}_{-y^{-1}q^{n}}T_M^{\ast}
\bigotimes_{n\geq 1}^{\infty}S_{q^n}T_{M}
\bigotimes_{n\geq 1}^{\infty}S_{q^n}T_M^{\ast}\right)
\mbox{,}
\eqno{(8)}
$$
and the integral (7) can be expanded as
$$
\chi(M;q,y)=\chi_y(T_M)+q\left\{\bigoplus_{s=0}^d\left[(-y)^{s+1}
\chi\left({\bigwedge}^sT_M\otimes T_M\right)\right.\right.
$$
$$
\left.\left.
+(-y)^{s-1}\chi\left({\bigwedge}^s
T_M\otimes T_M^{\ast}\right)
+(-y)^s\chi\left({\bigwedge}^sT_M\otimes (T_M\otimes T_M^{\ast})
\right)\right]\right\}+{\cal O}(q^{3/2})
\mbox{,}
\eqno{(9)}
$$
where we used formula (7). Note that the elliptic genus was independently
introdiuced in \cite{alva,alva1}.
For $q=0$ the elliptic genus reduces to a weighted sum over the 
Hodge numbers, namely 
$\chi(M;0,y)=\sum_{p,q}(-1)^{p+q}y^{p-\frac{d}{2}}h^{p,q}(M)$.

If $\chi({\cal H}_{(n)}^{{\Bbb Z}_n};q,y)$ admits the extension 
$\chi({\cal H};q,y)=\sum_{m\geq 0,\ell}C(nm,\ell)q^my^{\ell}$, the following 
result holds (see Ref. \cite{dijk97-185-197,dijk97-484-543}):
$$
\sum_{N\geq 0}p^N\chi(S^N{\cal H}_{(n)}^{{\Bbb Z}_n};q,y)=\prod_{m\geq 0,\ell}
\left(1-pq^my^{\ell}\right)^{-C(nm,\ell)}
\mbox{,}
\eqno{(10)}
$$
$$
\sum_{N\geq 0}p^N\chi(S^NM;q,y)=\prod_{n>0,m\geq 0,\ell}
\left(1-p^nq^my^{\ell}\right)^{-C(nm,\ell)}
\mbox{.}
\eqno{(11)}
$$
Here $p={\bf e}[\rho]$, $\rho$ and $\tau$ determine the complex K\"{a}hler
form and complex structure modulos of $T^2$ respectively, and $z$ parametries
the $U(1)$ bundle on $T^2$. 
The logarithm of the partition function $Z(p;q,y)$ is the one-loop free 
energy $F(p;q,y)$ for a string on $T^2\otimes M$:
$$ 
F(p;q,y)=\mbox{log}Z(p;q,y)=-\sum_{n>0,m,\ell}C(nm,\ell)\mbox{log}
\left(1-p^nq^my^{\ell}\right)
$$
$$
=\sum_{n>0,m,\ell,k>0}\frac{1}{k}C(nm,\ell)p^{kn}q^{km}y^{k\ell}
=\sum_{N>0}p^N\sum_{kn=N}\frac{1}{k}\sum_{m,\ell}C(nm,\ell)q^{km}y^{k\ell}
\mbox{.}
\eqno{(12)}
$$
The free energy can be written as a sum of Hecke operators $T_N$ 
\cite{lang76} acting on the elliptic genus of $M$ 
\cite{dijk97-185-197,borc95-120-161,grit95u-06}:
$F(p;q,y)=\sum_{N>0}p^NT_N\chi(M;q,y)$.

{\bf II}.\,\,\,\,\, The goal now is to calculate an asymptotic expansion
of the elliptic genus $\chi(S^NM;q,y)$. 
The Laurent inversion formula has the form
$$
\chi(S^NM;q,y)=\frac{1}{2\pi i}\oint\frac{Z(p,q,y)}{p^{N+1}}dp
\mbox{,}
\eqno{(13)}
$$
where the contour integral is taken on a small circle about the origin.
Let the Dirichlet series
$$
{\goth D}(s;\tau,z)=\sum_{(n,m,\ell)>0}\sum_{k=1}^{\infty}\frac{{\bf e}
[\tau mk+z\ell k]C(nm,\ell)}{n^sk^{s+1}}
\mbox{}
\eqno{(14)}
$$
converges for $0<\Re\,s<\alpha$. We assume 
that series (14) can be analytically continued in the region 
$\Re\,s\geq-C_0\,\,(0<C_0<1)$ where it is analytic except a pole of order 
one at $s=0$ and $s=\alpha$ with residue $\mbox{Res}[{\goth D}(0;\tau,z)]$
and $\mbox{Res}[{\goth D}(\alpha;\tau,z)]$ respectively. Besides 
let ${\goth D}(s;\tau,z)={\cal O}(|\Im\,s|^{C_1})$ uniformly in 
$\Re\,s\geq-C_0$ as $|\Im\,s|\rightarrow\infty$, where $C_1$ is a fixed 
positive real number.
If $t\equiv 2\pi(\Im\rho-i\Re\rho)$ then the Mellin-Barnes representation of 
the function $F(t;\tau,z)$ gives
$$
{\goth M}[F](t;\tau,z)=\frac{1}{2\pi i}\int_{\Re\,s=1+\alpha}t^{-s}\Gamma(s)
{\goth D}(s;\tau,z)ds
\mbox{.}
\eqno{(15)}
$$
The integrand in Eq. (17) has a first order pole at $s=\alpha$ and a second
order pole at $s=0$. Shifting the vertical contour from $\Re\,s=1+\alpha$ to
$\Re\,s=-C_0$ (this procedure is permissible) and making use of the residues
theorem one obtains

$$
F(t;\tau,z)=t^{-\alpha}\Gamma(\alpha){\rm Res}[{\goth D}(\alpha;\tau,z)]
+\lim_{s\rightarrow 0}\frac{d}{ds}[s{\goth D}(s;\tau,z)] 
$$
$$
-(\gamma+{\rm log}\,t){\rm Res}[{\goth D}(0;\tau,z)]
+\frac{1}{2\pi i}\int_{\Re\,s=-C_0}t^{-s}\Gamma(s){\goth D}(s;\tau,z)ds
\mbox{.}
\eqno{(16)}
$$ 
The absolute value of the integral in (16) can be estimated
to behave as ${\cal O}\left((2\pi\Im\,\rho)\right)^{C_0}$. 

We are ready now to state the result, which follows in a similar way 
from the Meinardus main theorem \cite{mein54-59-338,mein54-61-289,andr76}.
In the half-plane $\Re t>0$ there exists an asymptotic expansion for
$Z(t;\tau,z)$ uniformly in $|\Re\rho|$ for $|\Im\rho|\rightarrow 0, \,\,
|\mbox{arg}(2\pi i\rho)|\leq \pi/4,\,\,|\Re\rho|\leq 1/2$ and given by 
$$
Z(t;\tau,z)={\bf e}\left[\frac{1}{2\pi i}\left\{
\mbox{Res}[{\goth D}(\alpha;\tau,z)]\Gamma(\alpha)t^{-\alpha}
-\mbox{Res}[{\goth D}(0;\tau,z)]{\rm {log}}t \right.\right.
$$
$$
\left.\left.
-\gamma\mbox{Res}[{\goth D}(0;\tau,z)]
+\lim_{s\rightarrow 0}\frac{d}{ds}[s{\goth D}(s;\tau,z)]+{\cal O}\left(
|2\pi\Im\tau|^{C_0}\right)\right\}\right]
\mbox{.}
\eqno{(17)}
$$
The asymptotic expansion for the elliptic genus has the form
$$
\chi(S^NM;\tau,z)_{N\rightarrow \infty}={\cal C}(\alpha;\tau,z)N^{(2{\rm Res}
[{\goth D}(0;\tau,z)]-2-\alpha)/(2(1+\alpha))}
$$
$$
\times
{\bf e}\left[\frac{1+\alpha}{2\pi i\alpha}\left(\mbox{Res}
[{\goth D}(\alpha;\tau,z)]\Gamma(1+\alpha)\right)^
{1/(1+\alpha)}N^{\alpha/(1+\alpha)}\right]\left[1+{\cal O}(N^{-k})\right]
\mbox{,}
\eqno{(18)}
$$
$$
{\cal C}(\alpha;\tau,z)=\left\{\mbox{Res}[{\goth D}(\alpha;\tau,z)]
\Gamma(1+\alpha)\right\}
^{(1-2{\rm Res}[{\goth D}(0;q,y)])/(2+2\alpha))}
$$
$$
\times
{\bf e}\left[\frac{1}{2\pi i}\left(\lim_{s\rightarrow 0}\frac{d}{ds}
[s{\goth D}(0;\tau,z)]-\gamma\mbox{Res}[{\goth D}(0;\tau,z)]\right)\right]
\left[2\pi(1+\alpha)\right]^{1/2}
\mbox{,}
\eqno{(19)}
$$
where $k$ is a positive constant (see for detail Refs. \cite{andr76,
eliz94,byts96-266-1}). In above formulae the complete form of the prefactor
${\cal C}(\alpha;\tau,z)$ appears. The result (18), (19) has an universal 
character for all elliptic genera associated to Calabi-Yau manifolds.

Finally we go into some facts related to orbifoldized elliptic genus of $N=2$
superconformal field theory. The contribution of the 
untwisted sector to the orbifoldized  elliptic genus is the 
function $\chi(M;\tau,z)\equiv\phi(\tau,z)\equiv 
\sector00(\tau,z)$, whereas

$$
\phi\left(\frac{a\tau+b}{c\tau+d},\frac{z}{c\tau+d}\right)=
\sector00(\tau,z){\bf e}\left[\frac{rcz^2}{c\tau+d}\right],
\,\,\,\left( \begin{array}{ll}
a\,\,\,b\\
c\,\,\,d\\
\end{array}\right)\in SL(2,{\Bbb Z})
\mbox{,}
\eqno{(20)}
$$
$r=d/2$. The contribution of the twisted $\mu-$ sector projected by
$\nu$ is \cite{kawa94-414-191}:

$$
\sector{\nu}{\mu}(\tau,z)=\phi(\tau,z+\mu\tau+\nu){\bf e}
\left[\frac{d}{2}(\mu\nu+\mu^2\tau+2\mu z)\right]
,\,\,\,\,\,\,\,\mu,\nu\in {\Bbb Z}
\mbox{.}
\eqno{(21)}
$$
The orbifoldized elliptic genus can be defined by 
$$
\phi(\tau,z)_{\rm orb}\stackrel{def}=\frac{1}{h}\sum_{\mu,\nu=0}^{h-1}(-1)^
{P(\mu+\nu+\mu\nu)}\sector{\nu}{\mu}(\tau,z)
\mbox{,}
\eqno{(22)}
$$
where $P, h$ are some integers. Using transformation properties (21) in Eqs.
(18), (19) one can obtain the asymptotic expansion for the 
orbifoldized elliptic genus. In fact we can introduce a procedure, starting
with the expansion of the elliptic genus of the untwisted sector, to compute
the asymptotics of the elliptic genus of the twisted sector.

{\bf III}.\,\,\,\,\, Let $T$ and $U$ are the K\"{a}hler and complex structure 
moduli of the torus ${\bf T}^2$. 
The Narain duality group $SO(2,3;\Bbb Z)$ is isomorphic to Siegel
modular group $Sp(4,\Bbb Z)$ and it is convenient to combine the parameters
$T, U$ and a Wilson line modules $V$ 
into $2\otimes 2$ matrix belonging to Siegel upper half plane of genus $2$
$$
\Omega=\left( \begin{array}{ll}
T\,\,\,V\\
V\,\,\,U\\
\end{array}\right),\,\,\, {\Im T},\,\,{\Im U}>0,\,\,\,\,
\mbox{det}({\Im \Omega})>0
\mbox{.}
\eqno{(23)}
$$
One can use $\Theta$-transform \cite{kawa95u-46,neum96u-29}, which
converts the $K3$ elliptic genus $\chi(K3;q,y)=\sum_{m,\ell}C(4m-{\ell}^2)
{\bf e}[m\tau+\ell z]$ to the automorphic form \cite{grit95u-06,grit95u-08,
grit96u-22,grit96u-28,grit96u-02}
$$
\Phi(T,U,V)={\bf e}[T+U+V]\prod_{(n,m,\ell)>0}\left(1-{\bf e}[nT+mU+\ell V]
\right)^{C(4nm-{\ell}^2)}
\mbox{.}
\eqno{(24)}
$$
This form is the denominator product for the GKM based on the Cartan matrix
$$
{\cal A}=\left(\begin{array}{ll}
\,\,\,\,\,2\,\,\,\,\,-2\,\,\,\,-2\\
-2\,\,\,\,\,\,\,\,\,\,\,\,2\,\,\,\,\,-2\\
-2\,\,\,\,\,-2\,\,\,\,\,\,\,\,\,\,\,\,2\\
\end{array}\right)
\mbox{.}
\eqno{(25)}
$$
A counting formula for the degeneracies of a black hole preserving 1/4 of the 
supersymmetry of the background $IIA/K3\otimes T^2$, or $\mbox{heterotic}/T^6$
is presented in \cite{dijk97-484-543}. The degeneracies for the supersymmetric
black holes with (electric, magnetic) charge 
vectors ${\bf q}=(q_e,q_m)\in II^{6,22}\oplus II^{6,22}$  
have the form 
$$
d(q_e,q_m)=\oint \frac{{\bf e}\left[\frac{1}{2}{\bf q}\cdot\Omega
\cdot{\bf q}^t\right]}
{\Phi(T,U,V)}dTdUdV
\mbox{.}
\eqno{(26)}
$$

Let us expand the integrand of Eq. (26) as a formal Fourier series expansion
$$
{\bf e}\left[\frac{{\cal F}(\Omega)}{2\pi i}\right]\equiv\frac{{\bf e}[T+U+V]}
{\Phi(\Omega)}=
\sum_{(n,m,\ell)}{\bf D}(n,m,\ell){\bf e}[-(nT+mU+\ell V)]
\mbox{,}
\eqno{(27)}
$$
where $(n,m,\ell)$ are integers. Let
$$
{\goth D}_1(s;n,m)=\sum_{\ell}\frac{C(nm,\ell)}{{\ell}^s},\,\,\,
{\goth D}_1(s;n)=\sum_m\frac{\mbox{Res}[{\goth D}_1(\alpha_3;n,m)]}{m^s},
$$
$$
{\goth D}_1(s)=\sum_{n}\frac{\mbox{Res}[{\goth D}_1(\alpha_2;n)]}{n^s}
\mbox{}
\eqno{(28)}
$$
be the Dirichlet series which converges for $0<\Re \,s<\alpha_j\,\,(j=1,2,3)$.
We assume that the series ${\goth D}_1(s;n,m), {\goth D}_1(s;n)$ and 
${\goth D}_1(s)$ have poles of order
one at $s=\alpha_3, \alpha_2$ and $\alpha_1$ respectively. We assume also that
for $r\in \Bbb Z_{+}$ the series
$$
{\goth D}_2(s;-r)=\sum_n\frac{{\goth D}_1(-r;n)}{n^s},\,\,\,
{\goth D}_2(s;n,-r)=\sum_m\frac{{\goth D}_1(-r;n,m)}{m^s},\,\,\,
$$
$$
{\goth D}_2(s;\beta_2,-r)=\sum_n\frac{\mbox{Res}[{\goth D}_2
(\beta_2;n,-r)]}{n^s}
\mbox{}
\eqno{(29)}
$$
converges for $0<\Re\,s<\beta_j$ and have poles of order one at 
$s=\beta_1,\,\beta_2$ and $\beta_3$ respectively.

If ${\cal F}(\Omega)$ satisfy the expansion (27) and 
${\bf e}[-(T+U+V)]\equiv\exp[W+X+Y]$ then it can be shown that
$$
{\bf e}\left[\frac{{\cal F}(\Omega)}{2\pi i}\right]={\rm Res}[{\goth D}_1
(\alpha_1)]
\prod_{j=1}^3\Gamma(\alpha_j)
W^{-\alpha_1}X^{-\alpha_2}Y^{-\alpha_3}\zeta_R(\sigma)
$$
$$
+{\cal O}\left({\rm max}\left\{X^{-\alpha_2}Y^{\alpha_3},
W^{-\beta_1}Y^{-\alpha_3},W^{-\beta_3}X^{-\beta_2}\right\}\right)
\mbox{,}
\eqno{(30)}
$$
uniformly in $\left\{\Im W, \Im X, \Im Y\right\}$ as $\left\{\Re W,
\Re X,\Re Y\right\}\rightarrow 0$. Here $\zeta_R(\sigma)$ is the Riemann zeta 
function, $\sigma=1+\alpha_1+\alpha_2+\alpha_3$.
For $\{n,m,\ell \}\rightarrow\infty$ one has
$$
{\bf D}(n,m,\ell)\simeq{\cal I}(n,m,\ell){\bf e}
\left[\frac{\sigma -1}{2\pi i}\left({\rm Res}[{\goth D}_1
(\alpha_1)]\prod_{j=1}^3
\Gamma(\alpha_j)\zeta_R(\sigma) \right.\right.
$$
$$
\left.\left.
\times\left(\frac{n}{\alpha_1}\right)^{\alpha_1}
\left(\frac{m}{\alpha_2}\right)^{\alpha_2}
\left(\frac{\ell}{\alpha_3}\right)^{\alpha_3}
\right)^{\frac{1}{\sigma}}
\right]
\mbox{,}
\eqno{(31)}
$$
and
$$
{\cal I}(n,m,\ell)=\int_{{\cal G}}
{\bf e}\left[\frac{
{\cal F}(\Omega)+i(n\Im W+m\Im X+\ell \Im Y)}{2\pi i}\right]
\frac{d(\Im W)d(\Im X)d(\Im Y)}{(2\pi)^3}
\mbox{,}
\eqno{(32)}
$$
$$
{\cal G}=\left\{|\Re\,W|,\,\,|\Re\,X|,\,\,|\Re\,Y|<\epsilon<1\right\}
\mbox{.}
\eqno{(33)}
$$  
Using Eqs. (31) and (32) one can obtain
the degeneracy for a given electric and magnetic charge. The Eq. (31) gives 
an explicit counting formula of the asymptotic of microscopic black hole
states degeneracy and in particular allows one to analyze the asymptotic 
growth in more directions. This result is similar to the correct asymptotic 
growth as predicted from the macroscopic entropy formulas of extremal dyonic
black holes. 

There are intriguing relations between string duality, the theory of
generalized Kac-Moody algebras, and their associated automorphic forms. For
example, threshold corrections involve automorphic forms similar to those
associated to Kac-Moody algebras; black hole BPS states are counted by a
denominator product for a Kac-Moody algebra. In this note we used some of 
these relations in order to give the correct asymptotic expansion for the
degeneracies. We hope that proposed calculation will be interesting in view 
of future applications to concrete problems in string theory and in black
hole physics.  

\vspace{0.8cm}

{\bf Acknowledgements}
\vspace{0.3cm}

A.A. Bytsenko wishes to thank CNPq and the Department of Physics of
Londrina University for financial support and kind hospitality. The research
of A.A. Bytsenko was supported in part by Russian Foundation for Basic
Research (grant No. 98-02-18380-a) and by GRACENAS (grant No. 6-18-1997).
The work of S.D. Odintsov has been supported in part by Universidad del Valle 
(Colombia) and GRACENAS (grant No. 6-18-1997).

\end{document}